\documentstyle[fleqn,psfig]{tp}

\def\kms{\mbox{km s$^{-1}$}}

\def\ap3m{\mbox{AP$^3$M}}

\def\kms{\mbox{km/s}}

\def\kpch{\mbox{$h^{-1}$kpc}}

\def\LCDM{\mbox{\char'3CDM}}

\def\lcdm30art{\mbox{\char'3CDM$_{30}^{\rm ART}$~}}

\def\mpch{\mbox{$h^{-1}$ {\rm Mpc}}}

\def\sig8{\mbox{$\sigma_8$}}

\def\xidm{\mbox{$\xi_{dm}$}}
\def\xihh{\mbox{$\xi_{hh}$}}

\def\lesssim{\mbox{$_ <\atop{^\sim}$}}
\def\gtrsim{\mbox{$_ >\atop{^\sim}$}}

\begin{document}

\twocolumn[
\title{Evolution of halo-halo clustering and bias in a $\Lambda$CDM model}
\author{A.V. Kravtsov$^1$, A.A. Klypin$^1$, P.
  Col\'{\i}n$^2$, A.M. Khokhlov$^3$\\[3mm]
{\it $^1$Astronomy Department, NMSU, Dept.4500, Las Cruces, NM 88003-0001}\\
{\it $^2$Instituto de Astronom\'{\i}a, UNAM, C.P. 04510, M\'exico,
  D.F., M\'exico}\\
{\it $^3$ Laboratory for Computational Physics and Fluid Dynamics,}\\ 
{\it Code 6404, Naval Research Laboratory, Washington, DC 20375}}
\vspace*{16pt}   

ABSTRACT.\ We study the evolution of the halo-halo correlation function
and bias in a {\LCDM} model using very high-resolution $N$-body
simulations with dynamical range of $\sim 32,000$ (force resolution of
$\approx 2\kpch$ and particle mass of $\approx 10^9h^{-1} {\rm
  M_{\odot}}$).  The high force and mass resolution allows dark matter
(DM) halos to survive in the tidal fields in high-density regions and
thus prevents the ambiguities related with the ``overmerging problem.''
Numbers of galaxy-size halos in cluster-like objects in our
simulation are similar to the numbers of galaxies observed in real
clusters.  This allows us to estimate for the first time the evolution
of the correlation function and bias at small (down to $\sim 100\kpch$)
scales. We compare particle distribution, dark matter correlation
function, density profiles, and halo mass function produced with our
$N$-body code and corresponding results of the {\ap3m} simulations.

We find that at all epochs the 2-point correlation function of
galaxy-size halos $\xihh$ is well approximated by a power-law with
slope $\approx 1.6-1.8$. The difference between the shape of $\xihh$
and the shape of the correlation function of matter results in the {\em
  scale-dependent bias} at scales $\lesssim 7\mpch$, which we find to
be a generic prediction of the hierarchical models, independent of the
epoch and of the model details. We find that our results agree well with
existing clustering data at different redshifts, indicating the general
success of the picture of structure formation in which galaxies form
inside the host DM halos.  Particularly, we find an excellent agreement
in both slope and the amplitude between $\xihh(z=0)$ in our
simulation and the galaxy correlation function measured using the APM
galaxy survey. At high redshifts, the observed clustering of the
Lyman-break galaxies is also reasonably well reproduced by the models.

\endabstract]

\markboth{Kravtsov et al.}{Evolution of bias}

\small

\begin{figure}
 \centerline{\psfig{figure=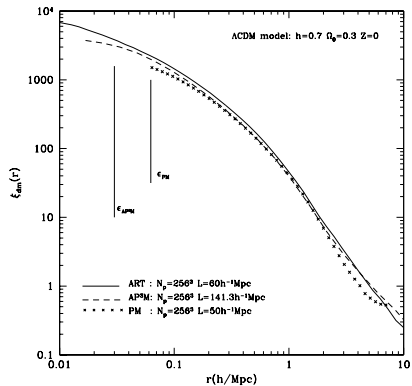,height=6cm}}
\caption[]{The comparison of the
  correlation functions of the dark matter {\xidm} in the \LCDM~ model
  estimated by different authors with different numerical resolutions
  and codes.  The {\em solid curve} shows {\xidm} in our {\LCDM} run,
  simulated using the ART code.  The {\em dashed curve} shows {\xidm}
  estimated by Jenkins et al. (1998, ApJ, 499, 20) using the \ap3m
  code.  The {\em crosses} show {\xidm} estimated by Klypin et al.
  (1996, ApJ, 466, 13) using the PM code.  The vertical lines
  indicate formal force resolution for each code (the line for the ART
  code at $1.8\kpch$ is off the plot).}
\label{fig:1}
\end{figure}

\section{Introduction}

It is widely believed that the distribution of galaxies is different from
the overall distribution of dark matter (DM). This difference, {\em the
  bias}, is crucial for comparisons between observations and
predictions of cosmological models.  Observations provide information
about the distribution of {\em objects} such as galaxies and galaxy
clusters. The models, however, most readily predict the evolution of the dark
matter distribution, which cannot be observed directly. The models,
therefore, should be able to predict the distribution of {\em objects} or,
conversely, predict how this distribution is different from that of
the dark matter (i.e., the bias). In the seminal paper, Kaiser (1984)
introduced the notion of bias to explain the strong clustering of
galaxy clusters. During the subsequent years, many analytical and
numerical studies have been done on the subject of the cluster and
galactic bias (see, e.g., references in Col\'{\i}n et
al. 1998). However, only recently the dynamic range of the pure $N$-body
simulations has become sufficiently high to allow to study unambigously
the {\em small-scale} bias of DM halos. In this contribution, we present the
results on the small-scale ($\approx 0.1-7\mpch$) halo-halo clustering and
bias of halos in one of the currently most successful cosmological
scenarios: flat CDM model with non-zero cosmological constant
($\Lambda$CDM). The parameters of the model are as follows:
$\Omega_0=1-\Omega_{\Lambda}=0.3$; $h=0.7$; $\sigma_8=1.0$; and age of
$\approx 13.4$ Gyrs. These results are a part of a larger study of the
bias evolution in different cosmological models described in Col\'{\i}n
et al. (1998) and we refer the reader to this paper for further
details. 

\section{Numerical Simulation}

One of the main goals of our study was to compute the correlation
function and the bias accounting for {\em all} DM halos, including
those inside groups and clusters. To assure that halos do survive in
clusters the force resolution should be $\sim 1-3h^{-1} {\rm kpc}$
(Moore et al. 1996; Klypin et al. 1998). Furthermore, if we aim
to study galaxy-size halos, the mass resolution should be $\lesssim
10^9h^{-1} {\rm M_{\odot}}$ to resolve galaxy-size halos ($M\gtrsim
10^{11}h^{-1} {\rm M_{\odot}}$) with at least $\approx 100$ particles.
The compromise between these considerations and the computational
expense determined the force and mass resolution of our simulation
carried out with the Adaptive Refinement Tree (ART) $N$-body code
(Kravtsov, Klypin \& Khokhlov 1997): particle mass of $\approx
10^9h^{-1} {\rm M_{\odot}}$ and peak spatial resolution of
$\eta/2\approx 2h^{-1} kpc$. The dynamical range $L_{box}/\eta$ of the
simulation implied by the force resolution is $\approx 16,000$
($32,000$ formal). The ART code integrates the equations of motion in
{\em comoving} coordinates.  However, its refinement strategy
is designed to effectively preserve the initial {\em physical}
resolution of the simulation. In order to prevent degradation of force
resolution in {\em physical} coordinates, the dynamic range between the
start and the end ($z=0$) of the simulation should increase by
$(1+z_i)$: i.e., for our simulations reach $512\times(1+z_i)=15,872$.
This is accomplished with the prompt successive refinements {\em in
  high-density regions} during the simulations. The peak resolution is
reached by creating a refinement hierarchy with six levels of
refinement. The spatial refinement is accompanied by the similar
refinement of integration time step.

\begin{figure*}
 \centerline{\psfig{figure=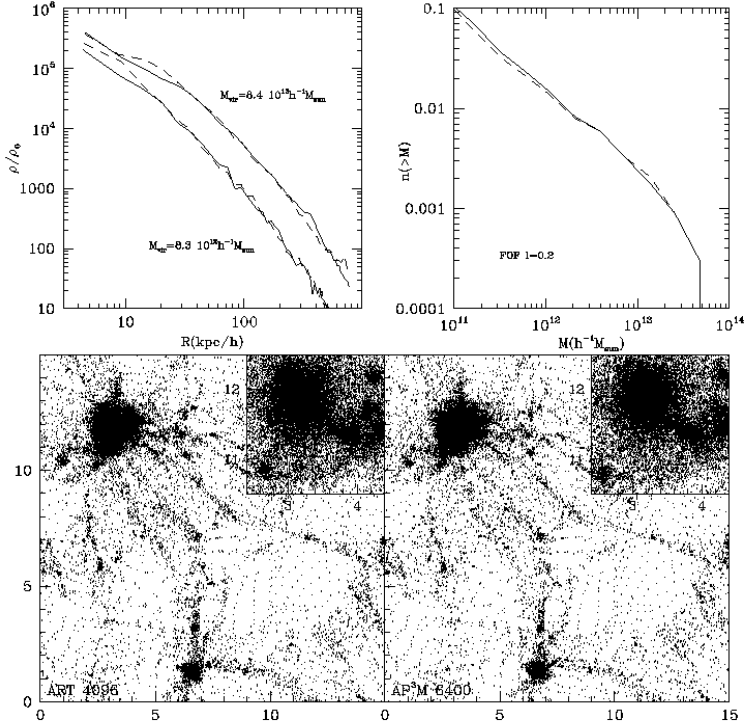,height=10cm}}
\caption[]{Comparison of $z=0$ outputs of ART and {\ap3m} CDM simulations
  started from the same initial conditions. Two bottom panels compare
  particle distributions in a slice through the simulation box. The
  upper panels compare halo density profiles (left) and halo mass
  functions (right).  The dashed (solid) lines show the ART ({\ap3m})
  results. }
\label{fig:2}
\end{figure*}

We use Bound Density Maxima (BDM, Klypin et al. 1998) algorithm to
identify galaxy-size halos that are isolated or belong to a larger
system.  The main idea of the BDM algorithm is to find positions of
local maxima in the density field smoothed at a certain scale and to
apply physically motivated criteria to test whether the identified site
corresponds to a gravitationally bound halo.  The detailed description
of the halo finder and simulation parameters can be found in Col\'{\i}n
et al. Here we present results of the tests of code performance based
on comparisons with {\ap3m} simulations.  In Figure 1 we compare the
2-point correlation function of the {\em dark matter} in our simulation
with similar estimates made with the PM and {\ap3m} codes (cf. fig.
caption).  We have made appropriate corrections to the amplitude of the
correlation function to account for slightly different normalizations
of the simulations.  Figure 1 shows that there is a very good agreement
between all estimates at scales $\approx (0.2-2)\mpch$.  The ART and
\ap3m estimates agree to better than 10\% at scales $0.03-7\mpch$!  On
larger scales the ART correlation function has a smaller amplitude than
that of the \ap3m due to a factor of 2.35 smaller box size of the
former.  This result is in agreement with previous studies which
concluded that the correlation function is underestimated at scales
$\gtrsim 0.1$ of the simulation box size.  Nevertheless, the agreement
is striking at smaller scales, given all the differences ({\em
  including cosmic variance}) between the estimates. We conclude,
therefore, that the correlation functions discussed in the next section
are robust at scales $\lesssim 7\mpch$.

The second test used smaller simulations of the CDM model to test the
distribution and properties of halos produced by the ART and {\ap3m}
codes. This test is a part of a comparison study carried out in
collaboration with Alexander Knebe and Stefan Gottl\"ober (AIP,
Potsdam). The ART simulations had a (formal) dynamic range of 4096,
while the {\ap3m} simulation had a dynamic range of 6400. Figure 2
shows comparison of the particle distributions, density profiles of
halos, and halo mass functions in the two simulations. Although the
small-scale details of the particle distribution are slightly different
(due to different time integration schemes used and somewhat different
spatial resolutions) the overall distribution is very similar in the
two runs. Almost identical results are also observed for the density
profiles and halo mass function.

\section{Results}

Our analysis shows that at all epochs correlation functions of halos
and dark matter are different in both shape and amplitude at scales
$<7h^{-1} {\rm Mpc}$. The difference in amplitude means that {\em halos
  are biased} with respect to dark matter, while the difference in
shape implies that the {\em bias is scale-dependent}.  Figure 3 shows
evolution of the bias with redshift at different scales and for halos
with different minimum circular velocities\footnote{The circular
  velocity is a mass indicator: the more massive is a halo, the higher
  is its circular velocity.}. We use a conventional definition of the
bias as a square root of the ratio of the halo and dark matter
correlation functions. The figure shows that bias undergoes a
significant evolution with redshift: it is high, $b\sim 2-4$ at high
redshifts $z\sim 3-5$, and then rapidly decreases at lower redshifts to
$b\sim 0.5-1$ at $z=0$. The high value of bias at early epochs is in
reasonably good agreement with the recent observations of strong galaxy
clustering at high redshifts (Steidel et al. 1998). The Fig. 3 also
shows that at high redshifts, bias depends on the mass of halos: the
massive halos are clustered more strongly than less massive ones. The
scale-dependency of the bias is illustrated in the bottom panel of the
Fig. 3, in which we show evolution of the bias at three different
scales for the halos with the circular velocity $> 100 {\rm km/s}$.
Although qualitatively the evolution is similar, the rate of evolution
and the value of bias are clearly dependent on the scale. In this panel
we also show analytical prediction of given by Moscardini et al. (1998)
for halos of mass $\geq 10^{11}h^{-1} {\rm M_{\odot}}$. There is a good
qualitative agreement between analytical predictions and our numerical
results.

\begin{figure}
\centerline{\psfig{figure=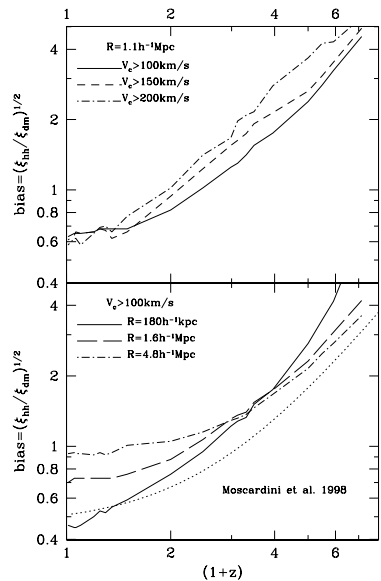,height=8cm}}
\caption[]{{\em Top panel:\/} The
  evolution of bias at {\em comoving} scale of $0.54\mpch$ for halos
  with different lower limit on the maximum circular velocity in the
  {\LCDM} simulation. The {\em bottom panel} shows dependence of
  the bias on ({\em comoving}) scale for halos with maximum circular
  velocity $>100\kms$.
}
\label{fig:3}
\end{figure}

We find that at all epochs the correlation function of DM halos is well
described by a power law $\xi(r)=(r/r_0)^{\gamma}$, where $r_0\approx
3-5h^{-1} {\rm Mpc}$ and $\gamma \approx -1.7$. This is in good
agreement with the shape and amplitude of the galaxy correlation
function at both low and high redshifts. Figure 4 shows the $z=0$
correlation functions of halos in our simulation and the correlation
function of galaxies from the APM galaxy survey. The figure shows that
there is an excellent agreement between the two at all scales probed by
our simulation. The dotted line in the bottom panel of Fig.4 shows the
correlation function of the dark matter. Note the difference in shape
between correlation functions of halos and dark matter. This difference
implies that the bias is scale-dependent; the bias as a function of
scale is shown in the upper panel of Fig.4. Note also that halo
correlation function has a lower amplitude than that of dark matter at
scales $< 5h^{-1} {\rm Mpc}$. This means that halos at these scales are
{\em anti-biased} with respect to the dark matter.

\begin{figure*}
 \centerline{\psfig{figure=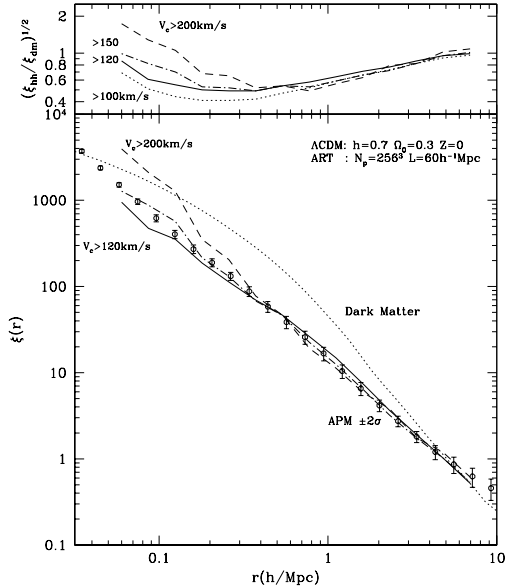,height=8cm}}
\caption[]{{\em Bottom panel:\/}
  Comparison of the halo correlation functions (solid, dot-dashed, and
  dashed curves) in the simulation with
  the correlation function of the APM galaxies (circles; Baugh 1996,
  MNRAS, 280, 267).
  The {\em dotted curve} shows the dark matter correlation function.
  {\em Top panel:} Dependence of bias on scale
  and maximum circular velocity. The curve labeling is the same as in
  the bottom panel, except that the {\em dotted line} now represents
  the bias of halos with $V_{max}>100\kms$.}
\label{fig:4}
\end{figure*}

Overall, the comparisons discussed above and more detailed analysis of
Col\'{\i}n et al. (1998) indicate that there is good agreement between
our results and the clustering data at both low and high redshifts.
This implies that hierarchical models in which observed galaxies form
in the host DM halos naturally explain the observed galaxy clustering
at different epochs, including excellent agreement with the accurately
measured $z=0$ correlation function. On the other hand, the generic
form of the bias evolution observed in the numerical simulations at
high redshifts agrees well with the prediction of the analytical models
based on the extended Press-Schechter formalism. This implies that we
understand the nature of the bias and the processes that drive its
evolution at high $z$. It is also reassuring that our results are in
good agreement with results of the studies which have used
semi-analytical modelling to follow evolution of galaxies in DM halos
(e.g., Kauffmann et al. 1998; Benson et al. 1998), as well as with the
results of direct numerical hydro simulations (Jenkins et al. 1998;
this meeting).  The success of the current theoretical models in
interpreting the clustering data forms a solid foundation for further
sophistication of the numerical models by including the processes important for
galaxy formation (such as dynamics of baryons, cooling, star formation,
and stellar feedback). We can expect, therefore, that the problem of
galaxy biasing will be tackled in the foreseeable future, which
would open for us great possibilities for the analysis of the ever
increasing amount of observations of galaxy clustering at low and high
redshifts.

\end{document}